\documentclass{aa}  

\usepackage[hang, flushmargin]{footmisc}

\usepackage{comment}
\usepackage{graphicx}
\usepackage{txfonts}
\usepackage{tikz,xcolor,hyperref}

\usepackage{soul}
\definecolor{jaune}{RGB}{255.0, 255.0, 0.0}

\hypersetup{
        colorlinks = true,
        urlcolor = blue,
        linkcolor = blue,
        citecolor=blue
}

\definecolor{lime}{HTML}{A6CE39}
\DeclareRobustCommand{\orcidicon}{%
        \begin{tikzpicture}
        \draw[lime, fill=lime] (0,0) 
        circle [radius=0.16] 
        node[white] {{\fontfamily{qag}\selectfont \tiny ID}};
        \draw[white, fill=white] (-0.0625,0.095) 
        circle [radius=0.007];
        \end{tikzpicture}
        \hspace{-2mm}
}

\foreach \x in {A, ..., Z}{%
\expandafter\xdef\csname orcid\x\endcsname{\noexpand\href{https://orcid.org/\csname orcidauthor\x\endcsname}{\noexpand\orcidicon}}
}

\begin{document}

    \title{The evolution of coronal shock wave properties and their relation with solar energetic particles}
    
    \titlerunning{Connecting SEP/coronal shock properties}

    \author{Manon Jarry\orcidA{}
    \inst{1}
    \and
    Nina Dresing\orcidB{}
    \inst{2}
    \and
    Alexis P. Rouillard\orcidD{}
    \inst{1}
    \and
    Illya Plotnikov\orcidE{}
    \inst{1}
    \and
    Rami Vainio\orcidC{}
    \inst{2}
    \and
    Christian Palmroos \orcidG{}
    \inst{2}
    \and
    Athanasios Kouloumvakos\orcidF{}
    \inst{3}
    \and
    Laura Vuorinen\orcidH{}
    \inst{2}
    }

   \institute{Institut de Recherche en Astrophysique et Planétologie (IRAP), CNRS, Université de Toulouse III-Paul Sabatier, France \newline e-mail: manon.jarry@irap.omp.eu
   \and
   Department of Physics and Astronomy, University of Turku, FI-20500 Turku, Finland
   \and
   The Johns Hopkins University Applied Physics Laboratory, Laurel, MD 20723, USA
   }
             
    \date{Received ...; accepted ...} 


\abstract
    {Shock waves driven by fast and wide coronal mass ejections (CMEs) are considered to be very efficient particle accelerators and are involved in the production of solar energetic particle (SEP) events. These events cause space weather phenomena by disturbing the near-Earth radiation environment. In past studies, we have analysed statistically the relation between the maximum intensity of energetic electrons and protons and the properties of coronal shocks inferred at the point of magnetic connectivity. The present study focuses on a gradual SEP event measured by STEREO-A and B on October 11, 2013. This event had the interesting properties that it (1) occurred in isolation with very low background particle intensities measured before the event, (2) was associated with a clear onset of SEPs measured in situ allowing detailed timing analyses, and (3) was associated with a fast CME event that was magnetically connected with STEREO-A and B. These three properties allowed us to investigate at a high cadence the temporal connection between the rapidly evolving shock properties and the SEPs measured in situ.}
    {The present study aims to investigate the relative roles of fundamental shock parameters such as compression ratio, Mach number and geometry, in the intensity and composition of the associated SEP event measured in situ.}
    {We use shock reconstruction techniques and multi-viewpoint imaging data obtained by STEREO-A and B, SOHO, and SDO spacecraft to determine the kinematic evolution of the expanding shock wave. We then exploit 3D magneto-hydrodynamic modelling to model the geometry and Mach number of the shock wave along an ensemble of magnetic field lines connected to STEREO-A and B, estimating also the uncertainties of the shock parameters. Using a velocity dispersion analysis of the available SEP data we time shift the SEP time series and analyse the relations between observed SEP properties and the modelled shock properties. We also study the energy dependence of these relations.}
    {We find a very good temporal agreement between the formation of the modelled shock wave and the estimated release times for both electrons and protons. The simultaneous release of protons and electrons suggests a common acceleration process. This early phase is marked at both STEREOs by elevated electron-to-proton ratios that coincide with the highly quasi-perpendicular phase of the shock. These findings suggest that the rapid evolution of the shock as it transits from the low to the high corona modifies the conditions under which particles are accelerated. We discuss these findings in terms of basic geometry and acceleration processes.}
    {}
    

\maketitle

\section{Introduction}
\label{sect_Introduction}

Fast coronal mass ejections (CMEs) drive shock waves in the solar corona capable of accelerating solar particles to high energies. The exact mechanisms involved in this acceleration process are still debated. In particular, the relative role of solar flares and shocks in the overall energisation process is still unclear. The advent of new techniques to derive the properties of shock waves in the solar corona based on multi-point imaging \citep{Kwon_2015, Rouillard_2016, Kouloumvakos_2019} has instigated new analyses on the relation between shock parameters and energetic particle events detected indirectly as electromagnetic radiation from the Sun \citep{Plotnikov_2017, Kouloumvakos_2020} or directly in situ \citep{Rouillard_2016, Lario_2016, Kouloumvakos_2019, Kouloumvakos_2023}.

The triangulation and modelling of expanding shocks provides important information on their rapidly evolving properties particularly during the early phase of CME formation and the onset of solar energetic particle (SEP) events \citep{Kwon_2015, Rouillard_2016}. This modelling reveals that the Mach number, compression ratio, and geometry are highly inhomogeneous over the shock surface and can change significantly in a matter of minutes as the shock expands in the highly structured solar atmosphere \citep{Rouillard_2016}. If the shock is indeed a prime contributor to the acceleration of particles, we should expect such varying shock properties to modify significantly the acceleration process \citep{Afanasiev_2018}.

Spacecraft measuring energetic particles and located at different positions in the inner heliosphere can be magnetically connected to very different regions of a shock's surface and potentially very different particle acceleration efficiencies. Modelling the early evolution of shock waves and how spacecraft connect magnetically to the shock is, therefore, necessary to compare SEP intensities profile with shock properties. 

Recent statistical studies have found high correlations between the shock Mach numbers and the maximum intensity of energetic solar protons and electrons observed during the events \citep{Kouloumvakos_2019, Dresing_2022}.
These studies have provided further support for a shock-related acceleration processes of protons and electrons such as diffusive shock acceleration and shock-drift acceleration as we will discuss in this paper.
These past studies compared the properties of shocks and SEPs but reduced to a single quantity deemed representative of the most intense aspect of each event, ignoring the time dependence of the events. Very few studies have followed with minute-cadence the early shock properties connected to a given probe and correlated both shock and SEP properties.
In the present paper, we focus on the first hours of a shock's expansion observed by multiple spacecraft and analyse further the relation between the rapidly evolving shock properties and SEP characteristics.
The advantage of the event analysed here is that it occurred in relative isolation with no contamination in energy bands of interest from other SEP events and powerful CMEs.
The SEPs were detected by the identical suites of particle detectors on both Solar-TErrestrial Relations Observatory spacecraft \citep[STEREO; ][]{Kaiser_2008} which provided high-quality measurements of protons, electrons and heavy ions.

This paper is structured as follows. In Sect. \ref{sect_event_presentation}, we present the 11 October 2013 event. In Sect. \ref{sect_shock_modelling} we present the methods used to reconstruct the 3D shape of the expanding pressure wave that eventually steepens into a shock wave, and model the properties of the latter. The analysis of individual shock parameters with SEP intensities is presented in Sect. \ref{sect_shock_SEP}. We interpret these results and discuss the possibility of other contributing factors to the acceleration process such as pre-event flaring activity in Sect. \ref{sect_interpretation}.

\section{The Oct 11, 2013 event}
\label{sect_event_presentation}

\begin{figure}[t]
    \centering
    \includegraphics[scale=0.34]{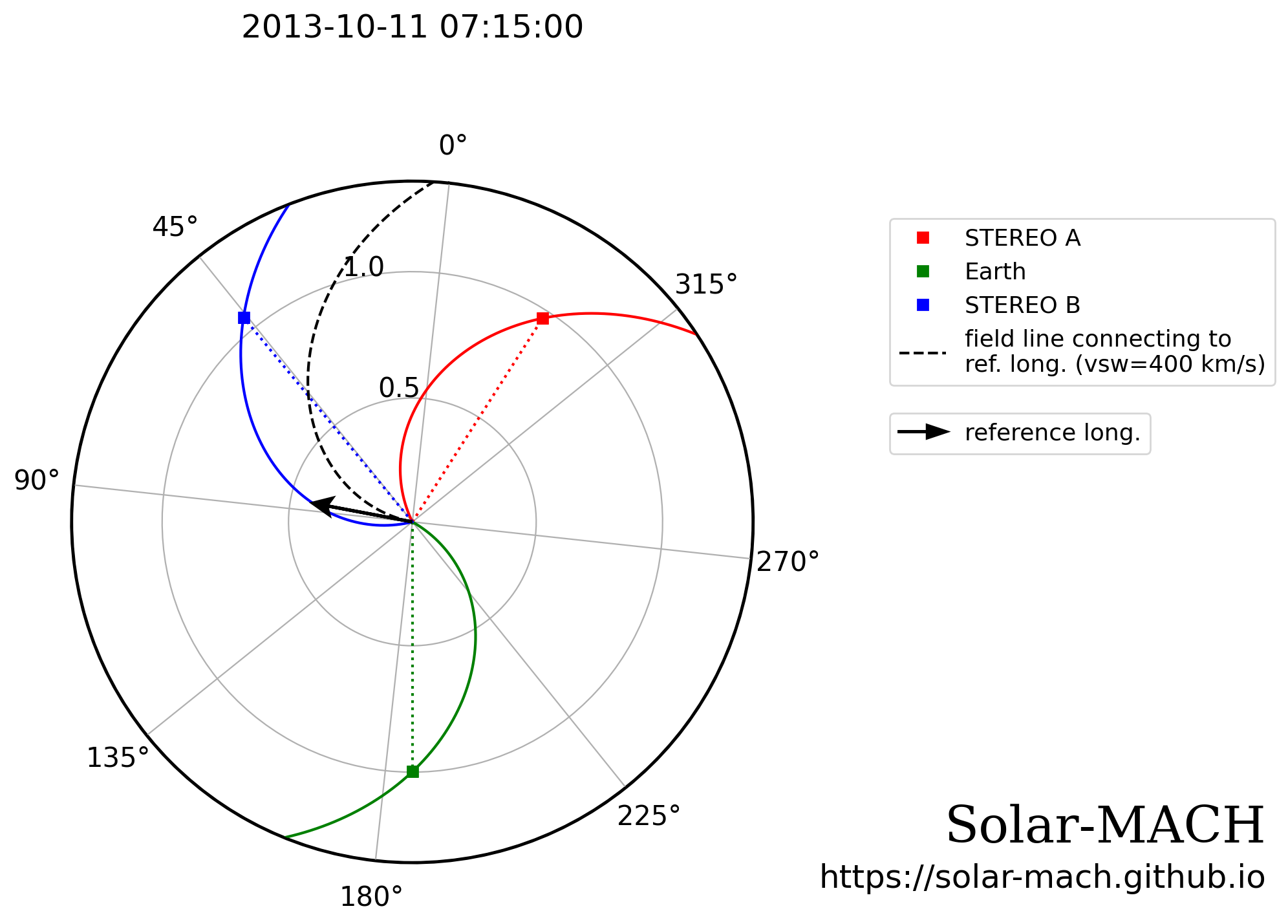}
        \caption{View of the ecliptic from the solar north that shows the spacecraft position around the Sun at the coronal mass ejection (CME) eruption time, i.e on 2013-10-11 07:15:00 UTC. The direction of propagation of the CME is shown by the black arrow, and corresponds to the latitude and longitude of the eruption in Carrington coordinates, (21$^{\circ}$, 103$^{\circ}$). This panel is produced using the Solar-MACH online tool \citep[\url{https://solar-mach.github.io/}, from][]{Gieseler_2023}.}
    \label{Solar_Mach}
\end{figure}

\begin{figure}[t]
    \centering
    \includegraphics[scale=0.45]{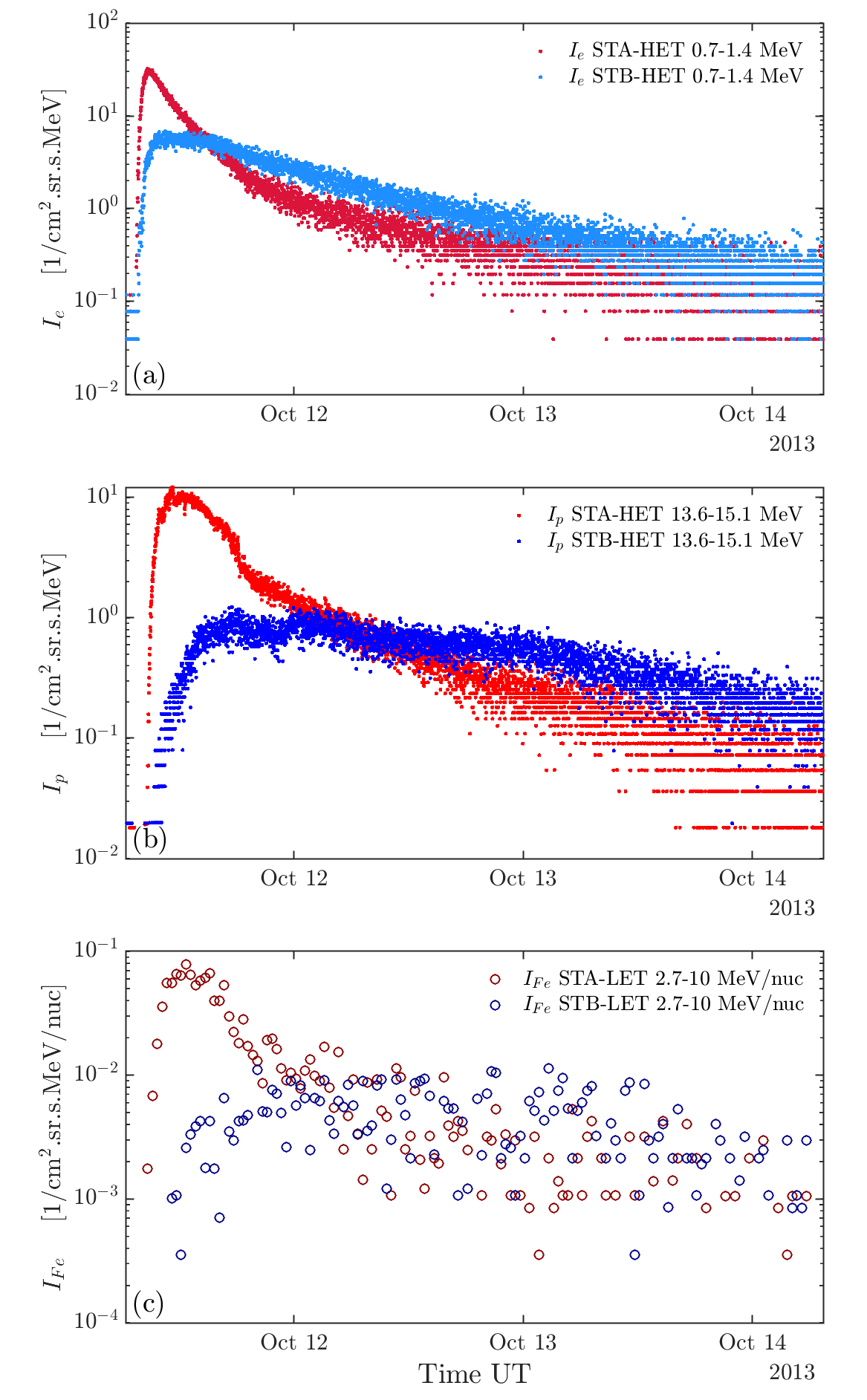}
        \caption{Panel (a) and (b): Energetic electron (panel a) and proton (panel b) measurements from HET detector onboard  STEREO-A (STA; red markers) and STEREO-B (STB; blue markers) from the 0.7-1.4 MeV energy channel for electrons and 13.6-15.1 MeV for protons.
        Panel (c): Energetic iron measurements (30min-averaged) from LET detector onboard STA and STB (respectively in dark red and dark blue) from the 2.7-10 MeV energy channel.}
    \label{STA_and_STB_SEP}
\end{figure}

\begin{figure*}[t]
    \centering
    \includegraphics[scale=0.42]{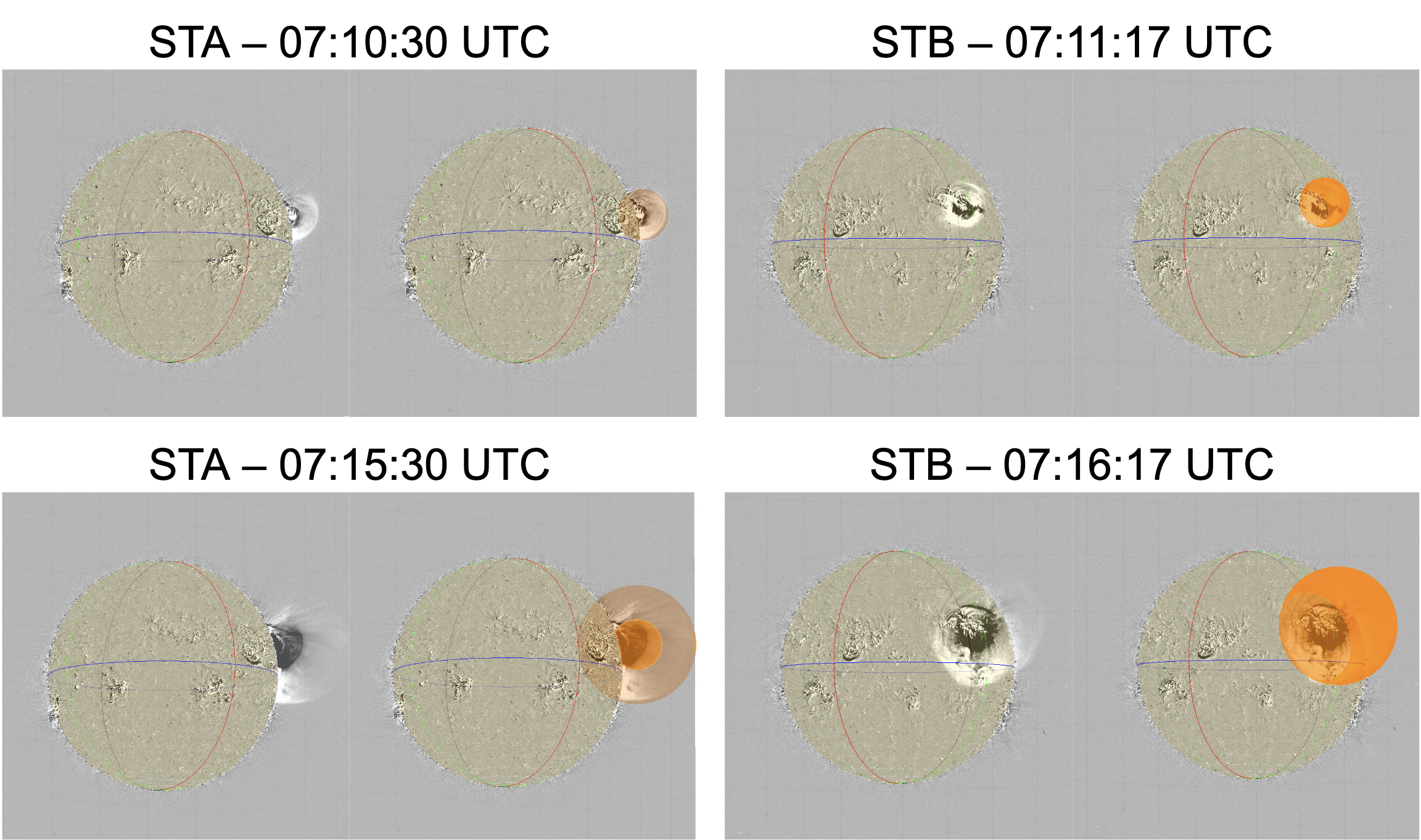}
        \caption{Running difference EUV images of the CME and the shock wave from two of the three viewpoints utilised in this study. Left panels are images from STEREO-A and right panels images from STEREO-B. The shape of the triangulated shock-wave using a 3D ellipsoid model is shown is orange, for t=07:11 UTC (top panels) and t=07:16 UTC (bottom panels).}
    \label{fig_shock_reconstruction}
\end{figure*}

On 2013 Oct 11 at 07:01 UT a flare occurred at an active region (AR) located at N21E106 position (in Stonyhurst coordinates) and was observed in soft X-rays by the Geostationary Operational Environmental Satellite Network (GOES) and the MErcury Surface, Space ENvironment, GEochemistry and Ranging \citep[MESSENGER;][]{Gold_2001}. The source location was 10 degrees behind the east limb as viewed from Earth at the time of the flare. This flare marked the onset of a powerful CME and a wealth of extreme ultra-violet (EUV), visible, radio, X-ray and $\gamma$-ray radiations observed by STEREO, the Solar and Heliospheric Observatory \citep[SoHO;][]{Domingo_1995} and the Solar Dynamics Observatory \citep[SDO;][]{Lemen_2012}, RHESSI, GOES, and Fermi spacecraft \citep[e.g., ][]{Pesce-Rollins_2015, Klassen_2016, Plotnikov_2017}.

\citet{Pesce-Rollins_2015} report that STB/EUVI recorded a flare starting at 07:01 and peaking at 07:25 associated with this event, situated $\approx$ 9.9 degrees behind the Earth east limb. They estimate the GOES class flare at M4.9 \citep{Nitta_2013}. According to \citet{Ackermann_2017}, \citet{Plotnikov_2017} and \citet{Share_2018} the flare occurred at an active region with coordinates N21~E103--106.

We chose this event because it exhibited a number of interesting properties that allowed a detailed comparison between the evolution of the shock and the SEPs.

First, the orbital configuration of the spacecraft, shown in Fig. \ref{Solar_Mach}, provides coronograph observations of three well separated viewpoints, which allowed an accurate fitting of the expanding shock wave. At the time, the longitudes and latitudes (lon, lat) of spacecraft around the Sun were (333.6$^\circ$, -7.3$^\circ$) for STEREO-A (STA), (45.8$^\circ$, -2.6$^\circ$) for STEREO-B (STB) and (186.2$^\circ$, 6.1$^\circ$) for the Earth, in Carrington coordinates.

Second, the SEP event occurred in isolation to previous solar activity and the two STEREO spacecraft provided high quality and continuous measurements of electrons, protons and ions throughout the event as shown in Fig. \ref{STA_and_STB_SEP}. Protons and electrons were recorded by the Solar Electron and Proton Telescope \citep[SEPT;][]{Muller-Mellin_2008} and the High Energy Telescope \citep[HET;][]{vonRosenvinge_2008}, respectively, on each STEREO spacecraft. The SEPT energy channels range from 84 to 6500 keV for protons with 30 channels, and from 45 to 425 keV for electrons with 15 channels. The HET energy channels range from 13.6 to 100 MeV for protons with 11 channels and from 0.7 to 4 MeV for electrons with 3 channels. This provides us a total coverage of 18 energy channels for electrons and 41 energy channels for protons in a very broad energy range. In addition, the low pre-event background (for STA/HET and STB/HET energy bands) and the clear onsets of the SEPs provided very good conditions to infer the release times of the different particles populations using a velocity dispersion analysis \citep{Vainio_2013}.

Third, as we will show in this paper the two STEREO spacecraft were magnetically connected to the Eastern and Western flanks of the shock during its formation in the low corona (see Fig. \ref{sect_shock_modelling}) allowing a detailed comparison between shock evolution and SEP production.

Although this SEP event is a gradual one, it exhibits a composition of heavier ions similar to impulsive SEP events. STEREO observations from the Low-Energy Telescope \citep[LET; ][]{Mewaldt_2008} show the presence of heavy ions and a high Fe/O ratio (>1 for energies >1 MeV/nuc), which are often observed during impulsive SEP events associated with solar flares. Such a composition can occur in gradual events as well \citep[e.g.][]{Cohen_1999} and may be related to a shock-related re-acceleration of pre-energised particles from a flare-related process \citep{Cohen_1999, Tylka_2005, Tylka_2006}. Furthermore, as we show in panel (a) and (b) of Fig. \ref{STA_and_STB_SEP}, STA recorded higher SEP intensities than STB, with a steeper rise phase, and in panel (c), higher Fe intensity during the first hours of the event.

\section{Modelling of the expanding pressure wave}
\label{sect_shock_modelling}

\begin{figure*}[t]
    \centering
    \includegraphics[scale=0.6]{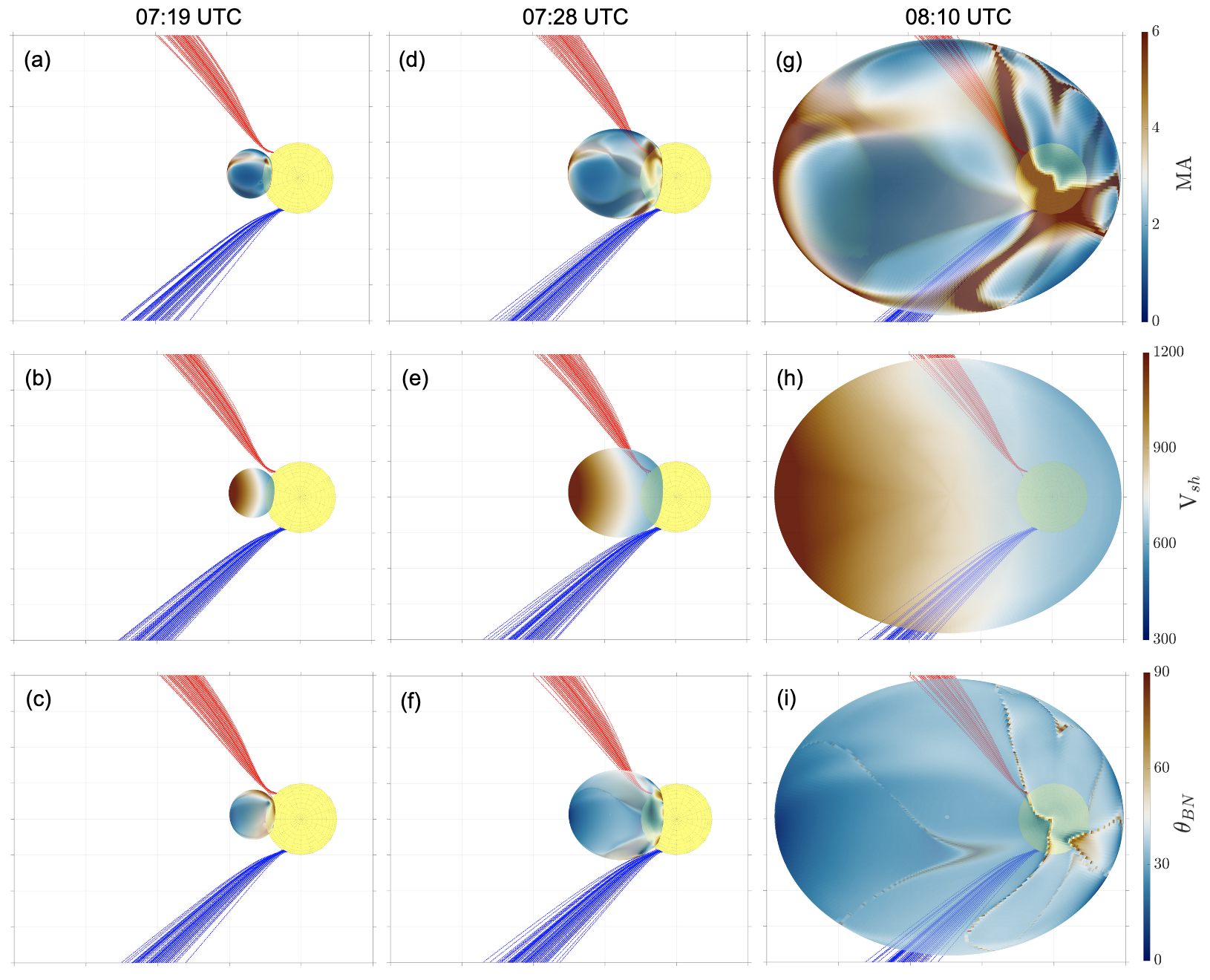}
        \caption{A view of the ecliptic plane as seen from north that show the results from the shock model in 3D and the magnetic connectivity. For each panel, the yellow sphere represents the Sun in true scale, the coloured ellipsoid represents the reconstructed pressure/shock wave with a colormap that corresponds to the different shock parameters, the red and blue lines represents the magnetic field lines connected to STA and STB spacecraft, respectively. These hundred connectivity lines are created according to a random number of latitude and longitude taken in a normal distribution around main latitude and longitude. The surface of the pressure/shock wave is coloured according to $M_A$, $V_{sh}$, and $\theta_{BN}$ values in the top, middle, and bottom rows, respectively. From left to right, the panels show the shock position at 07:19, i.e. just before its connection to STA field lines, 07:28 just before the connection to STB field lines, and 08:10 when the shock has been connected to both spacecraft (times in UTC).}
    \label{shock_3D_and_ST_FL}
\end{figure*}

To derive the magneto-hydrodynamic (MHD) properties of the shock and the magnetic connectivity of the shock to STA and STB, we followed the approach of \citet{Rouillard_2016}. This consists of a combination of image-based triangulation techniques to reconstruct the shock in 3D and 3D MHD modelling of the background corona, here provided by Predictive Sciences Inc. \citep[PSI; ][]{Lionello_2009}. This combination of techniques has been exploited in numerous studies, for instance in \citet{Kouloumvakos_2019} that produced a catalogue of 33 reconstructed coronal pressure waves over a complete solar cycle, using multi-viewpoint imaging data from near-Earth spacecraft and STEREO. This catalogue is currently maintained with the latest data provided by Solar Orbiter and Parker Solar Probe through the H2020 SERPENTINE project\footnote{\url{www.serpentine-h2020.eu}}. The catalogue was exploited by \citet{Jarry_2023} to extract the general geometric and kinematic properties of these modelled shock waves.

The shock-triangulation technique employed to study the present event in \citet{Kouloumvakos_2019} assumed an ellipsoidal shape with a derived central axis orientation aligned with a heliographic latitude of 12$^\circ$ and Carrington longitude of 86$^\circ$.  We carried out a more detailed triangulation of the first few minutes of the event that are of particular interest to the present study. Fig. \ref{fig_shock_reconstruction} shows the result of this initial shock phase reconstruction. The derived central axis of the ellipsoid is first aligned with a Carrington latitude of 16$^\circ$ and longitude of 84$^\circ$.

Fig. \ref{shock_3D_and_ST_FL} shows the reconstructed shock location at 07:19, 07:28 and 08:10 UTC and the magnetic field lines connected to STA and STB. The red and blue lines represent the modelled magnetic field lines connecting the spacecraft to the Sun's surface. Magnetic field lines connecting the spacecraft to the Sun's surface were obtained using a Parker spiral from the spacecraft to 20 solar radii, and then traced lower in the corona using MHD cubes provided by Predictive Science Inc. (PSI) (\url{https://www.predsci.com/mhdweb/home.php}). Here we utilise the thermodynamic Magneto-hydrodynamic Algorithm outside a Sphere (MAST) MHD modelling. This model uses for its inner boundary condition a magnetogram recorded by the Helioseismic and Magnetic Imager (HMI) onboard Solar Dynamic Observatory (SDO). The field line with nominal connectivity to each spacecraft is shown as a solid line. Then taking a random location following a normal (Gaussian) distribution of approximately 8 degrees around the latitude and longitude of the nominal magnetic field line at 20 $R_\odot$ we estimate another 99 field lines for each spacecraft. This corresponds to an difference in solar wind speed of approximately 25 $\pm$ 5 km/s.

\begin{figure}[t]
    \centering
    \includegraphics[scale=0.45]{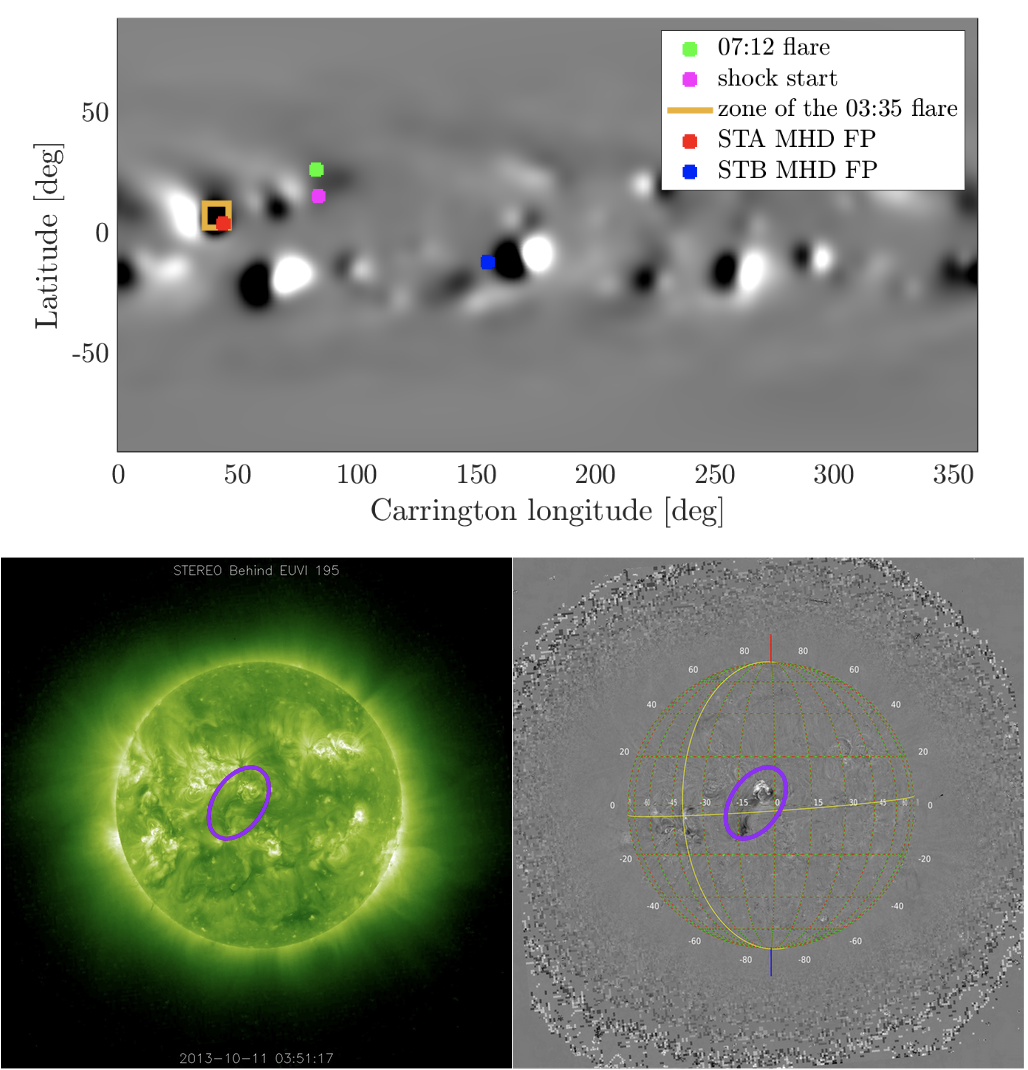}
        \caption{First panel: Representation of the Sun magnetogram of the Carrington rotation 2143, for MHD cube. The colour is given according to the magnetic field value from the cubes. The green dot indicates the location of the flare, whereas the pink dot indicates the projected position of the shock nose at its first reconstruction. The yellow square indicates the location with uncertainties of a flare occurring at 03:35 $\pm$ 00:05am which could have influenced the SEP event received by STA. The red and blue dots represents respectively the STA and STB footpoints obtained for MHD cubes. See the text for more information.
        Second panel: Image in EUVI 195 from STEREO-B, at 03:51:17 UTC (left) and integrated trough 03:00 UTC and 04:00 UTC (right). The purple ellipsoid indicates the zone of the flare occurring at 03:35 $\pm$ 00:05 UTC, corresponding to the yellow square in first panel.}
    \label{magnetogram}
\end{figure}

The top panel of Fig. \ref{magnetogram} presents the locations of connectivity footpoints, the location of the associated flare and the origin of the expanding pressure wave, and the location of a flare that occurred prior to the SEP event, at the magnetogram used for the MHD simulation. These connectivity points (blue markers for STB and red for STA) can be compared with the location that the shock wave originated (the magenta marker) and the main flare associated with the CME event (the green marker). We find that the footpoints of field lines connected to STA and STB were situated at Carrington longitudes 44.0 and 155.0 degrees, e.g. 40 degrees and 70 degrees away from the flaring site, respectively.

Another flare occurred some four hours before our main event and was well connected with STA as the magnetic connectivity analysis from the MHD simulation reveals. This flare occurred at 03:35 $\pm$ 00:05 UTC around (lon, lat) = $(41.3^\circ, 8^\circ)$ in Carrington coordinates (yellow square on magnetogram). Since this flare occurred on the far side of the Sun, not visible from Earth, there were no GOES X-ray measurements but it was visible in STB/EUVI 195 \AA ~ images (see Fig. \ref{magnetogram}, bottom panels). The flare had no type III radio burst counterpart in STEREO SWAVES \citep{Bougeret_2008} data and although being magnetically connected with STA, no SEPs were detected at the spacecraft during that flare or at any time until the main 2013, Oct 11 event. These observations suggest that if particles were energised by this pre-event flare they could not escape from the flaring region and remained confined in the magnetically closed corona. Nevertheless, as we shall argue later, these possibly pre-accelerated trapped particles could have been accelerated further by the shock to produce a more intense SEP event at STA.

A first important feature of this event is that the CME emerges from a region located between the two nominal connectivity points of STA and STB, a few degrees closer to the connection point of STA (see Fig. \ref{magnetogram}). This means that the modelled pressure wave intersected the field lines connected to STA and STB at around the same time with an estimated connection time to STA at 07:20 UT and to STB at 07:29 UT. We note that radio observation for this event show a drifting metric radio type II burst measured between 07:10 and 07:20 UT by the Culgoora spectrograph that corresponds to a shock speed of 924 km/s \citep{Plotnikov_2017}, which is close to the CME/shock speeds derived from the shock modelling during the same time interval. This suggests that at least some regions of the pressure wave have already developed into a shock wave by that time. This is also confirmed by the shock model that shows the presence of super-Alfvénic regions already at 07:11 UTC and quasi-perpendicular shock regions which are key properties for the production of type II emission \citep{Kouloumvakos_2021, Jebaraj_2021}.

Both spacecraft connect initially to the flank of the pressure wave very low in the corona. Fig. \ref{shock_3D_and_ST_FL} shows that the shock has already formed at the magnetic connection locations. This suggests that the acceleration process has already been initiated and this could explain the observed nearly simultaneous SEP onset times at the two spacecraft as seen in Fig. \ref{STA_and_STB_SEP}. Additionally, the shock geometry is initially highly quasi-perpendicular at the locations of the magnetic connection of the spacecraft to the shock.

Although the multi-point coronagraph observations of the three observing spacecraft were ideal (Fig. \ref{Solar_Mach}), our results contain some uncertainty, which stem from the quality of the 3D modelling of the shock wave, which is done manually from STEREO and SoHO coronograph images.

To evaluate the uncertainties associated to the magnetic field line reconstruction, in addition to the modelling of 100 magnetic field lines, we varied the tracing :
\begin{itemize}
\item by using Parker spirals connecting spacecrafts to the MHD cube at different solar radius : 30, 25, 20 and 15 R$_\odot$
\item by varying the solar wind speed of +/- 10, 20, 30 and 50 km/s compared to the measured value
\item by using the Potential Field Source Surface (PFSS) model instead of MHD, which use a Parker spiral until 2.5 solar radius and then PFSS
\end{itemize}

In all these configurations, STA footpoints are very stable because of the converging magnetic field lines whatever the inputs. They are localised on the solar surface in a radius of $\pm 15^{\circ}$ in latitude and longitude. We are therefore very confident about the values of the shock parameters and their uncertainties.

Along STB field lines however the reconstruction was more prone to uncertainties, with variations according to the inputs parameters for the field line construction. By increasing the wind speed by 50 km/s, for example, the MA values could even exceed those obtained for STA for some field lines. These uncertainties stemmed from the proximity of STB footpoints to the heliospheric current sheet \citep{Rouillard_2016}.\\

\begin{figure*}[t]
    \centering
    \includegraphics[scale=0.33]{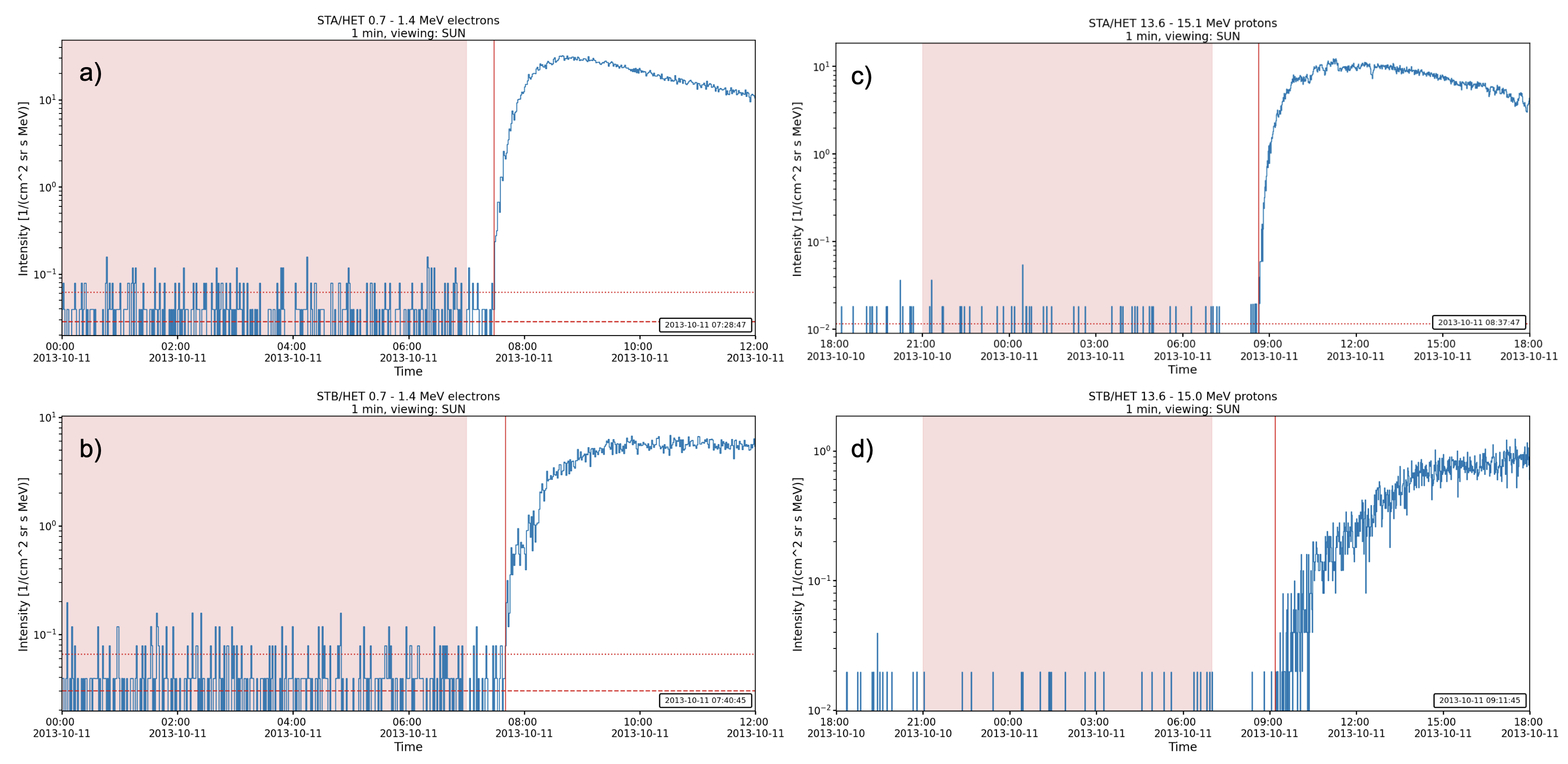}
        \caption{SEP intensities as a function of time, with onsets (vertical red line and box in the south east corner). Panel (a) and (b): 0.7-1.4 MeV electrons intensity, respectively for STA and STB.
        Panel (c) and (d): 13.6-15.1 MeV protons intensity, respectively for STA and STB.}
    \label{SEPs_onsets}
\end{figure*}

For the results in this article, we kept 20 R$_\odot$ for the connection between the Parker Spiral and the MHD cube. We chose to use the $V_{\mathrm{sw}}-50$ km/s configuration, justified by recent studies on the acceleration of the wind during its propagation \citep{Dakeyo_2022}. This choice leads to a higher Mach at STA than at STB, favouring the shock scenario. However, as we will discuss in Sect. \ref{sect_interpretation}, this does not seem to explain the higher particle flux received by STA.

\section{Deriving the solar release times of the SEPs}
\label{sect_analyse_particles}

\begin{figure}[t]
    \centering
    \includegraphics[scale=0.25]{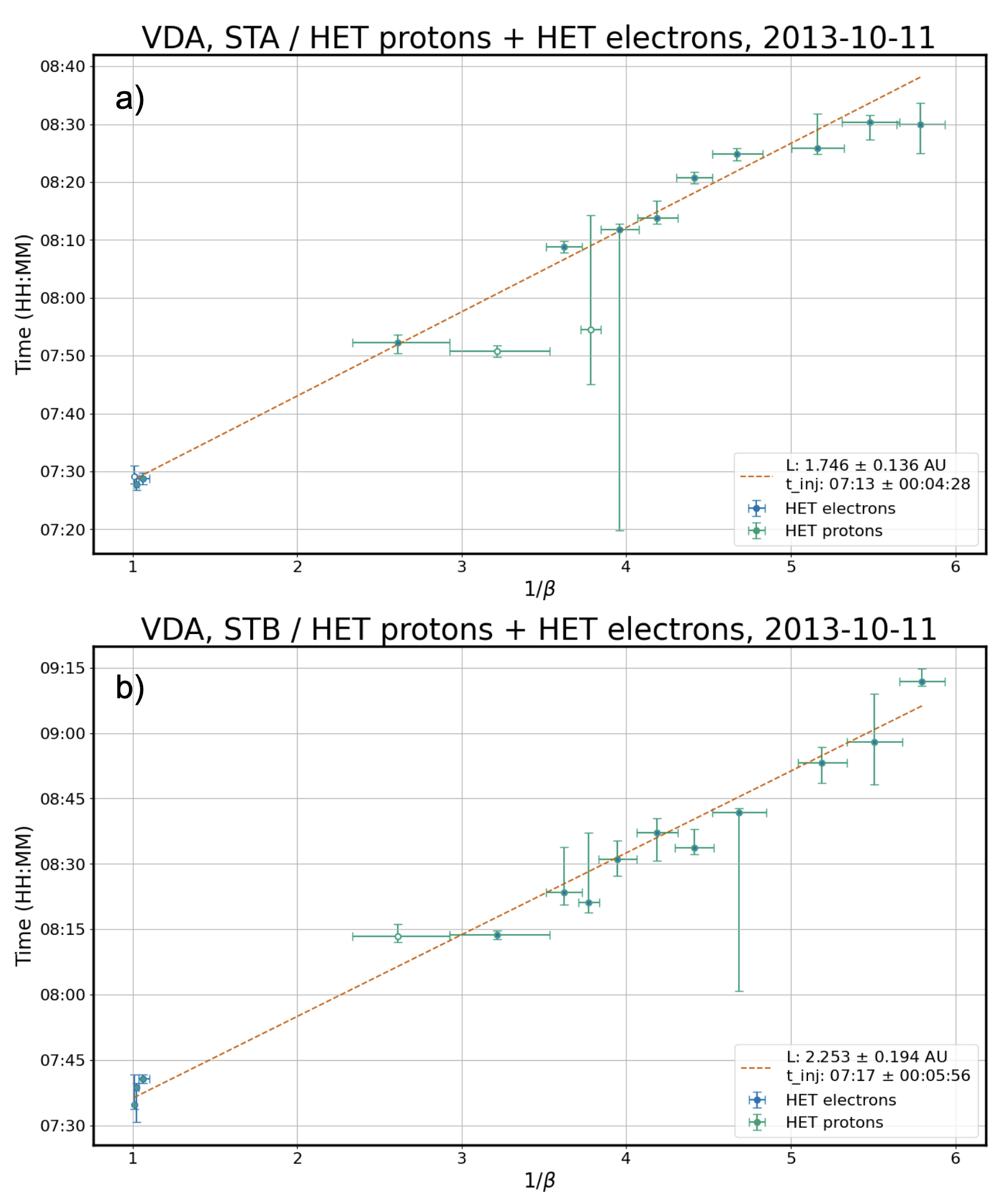}
        \caption{Velocity dispersion analysis (VDA) for the October 2013, 11 event. For each panel, observational time in HH:MM format as a function of $1/\beta$, the plasma parameter (see the text for more). Panel (a): VDA for STEREO-A. Panel (b): VDA for STEREO-B. Blue points correspond to electrons and green points to protons. Only filled circles are included in the fit (dashed red line).}
    \label{VDA}
\end{figure}

Since the aim of the study is to evaluate solar conditions around the release time of the SEPs, we performed a velocity dispersion analysis (VDA) to determine the solar release time and path-length travelled by the particles. For each proton and electron energy channel where the particle onset was clearly visible relative to the particle background, an onset time was determined using a Poisson-CUSUM method \citep{Palmroos_2022}. An example of the derived onset times using this method is presented in Fig.~\ref{SEPs_onsets}. As is it frequently done in VDA, we assume that protons and electrons are released at the same time and the VDA is then derived by identifying the onsets of particles in each energy channel. Fig. \ref{VDA} shows the two VDAs of STA (panel a) and STB (panel b), calculated for electrons and protons recorded by the HET instrument. Time is in hours on the event onset times are given as a function of $1/\beta = c / V$, where $V$ is the particle velocity and $c$ the speed of light.

A least-squares fit (represented by the red dashed line) is done using electron and proton points together (only filled symbols are included in the fit). The resulting injection time (Sun reference) is 07:13 $\pm$ 4.28 minutes UTC for STA and 07:17 $\pm$ 5.56 minutes UTC for STB. The associated calculated path length is 1.75 $\pm$ 0.14 AU for STA and 2.25 $\pm$ 0.19 AU for STB. These release times are then used to shift the time series of protons and electrons for each energy channel back in time so that they can be compared with a common "solar time" (ST).

A time shift of 8.33 minutes, corresponding to light travel time to the STEREO, is also subtracted from the time evolution of the shock parameters which were derived from STEREO imaging data as well as all timings of electromagnetic radiations. Based on this common ST, we can compare the SEP time series directly with the evolution of the shock parameters.

\section{Comparison between shock parameters and SEPs}
\label{sect_shock_SEP}

\begin{figure*}[h!]
    \centering
    \includegraphics[scale=0.36]{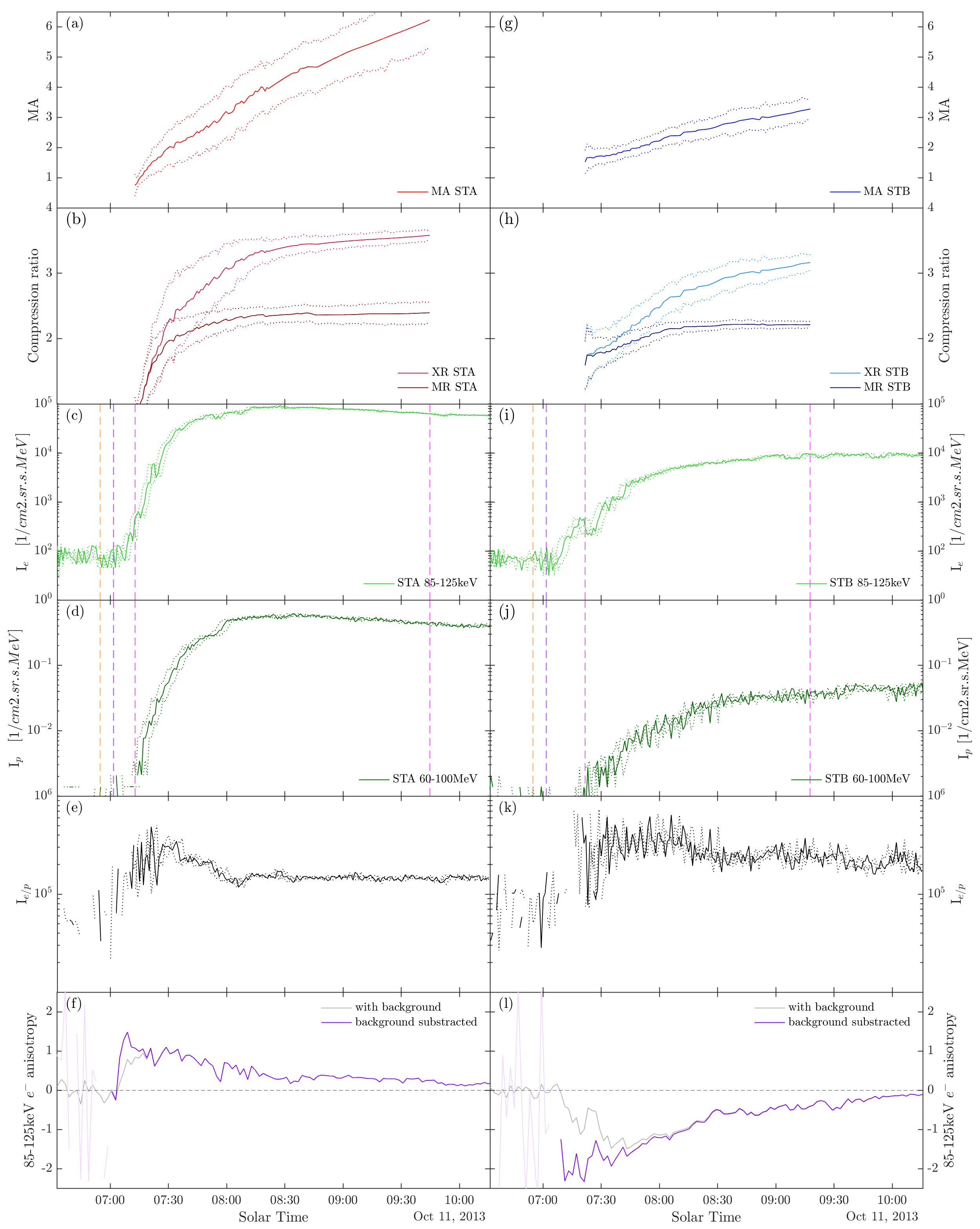}
        \caption{Time-series for STEREO-A (left panels) and STEREO-B (right panels) for the October 2013, 11 event.
        Panel a [g] the Alfvénic Mach Number ($M_A$) of the shock on the intersection point with the STA (in red) [STB (in blue)] field line. The solid line represents $M_A$ for the original magnetic field line connecting the surface of the Sun to STA [STB] whereas the dashed lines represent the uncertainties on the $M_A$. These are calculated using 100 magnetic field lines created by taking random (lon,lat) in a Gaussian around the original latitude and longitude of the magnetic field line at 20 solar radius (see Sect. \ref{sect_shock_modelling} for more details).
        Panel b [h] the density compression ratio (XR) and the magnetic compression ratio (MR) of the shock on the intersection point with the STA (in red) [STB (in blue)] field line. As for panel a [g], the solid line represents values for the original magnetic field line connecting the surface of the Sun to STA [STB] whereas the dashed lines represent the uncertainties.
        Panel c [i] the electrons received by STA [STB] SEPT in the 85-125 keV energy band.
        Panel d [j] protons received by STA [STB] HET in the 60-100 MeV energy band. The two vertical magenta dashed lines in panel c, d, i and j indicate the time of shock connection and the end time of the shock, while the orange and purple dashed lines represents respectively the timings of the hard X-ray and type II radio burst.
        Panel e [k] electron to proton ratio from the two precedent times-series for STA [STB].
        Panel f [l] the anisotropy (in grey) and the background-subtracted anisotropy (in purple) of STA [STB] SEPT electrons of 85-125 keV.}
    \label{MA_time-series}
\end{figure*}

\begin{figure*}[t]
    \centering
    \includegraphics[scale=0.6]{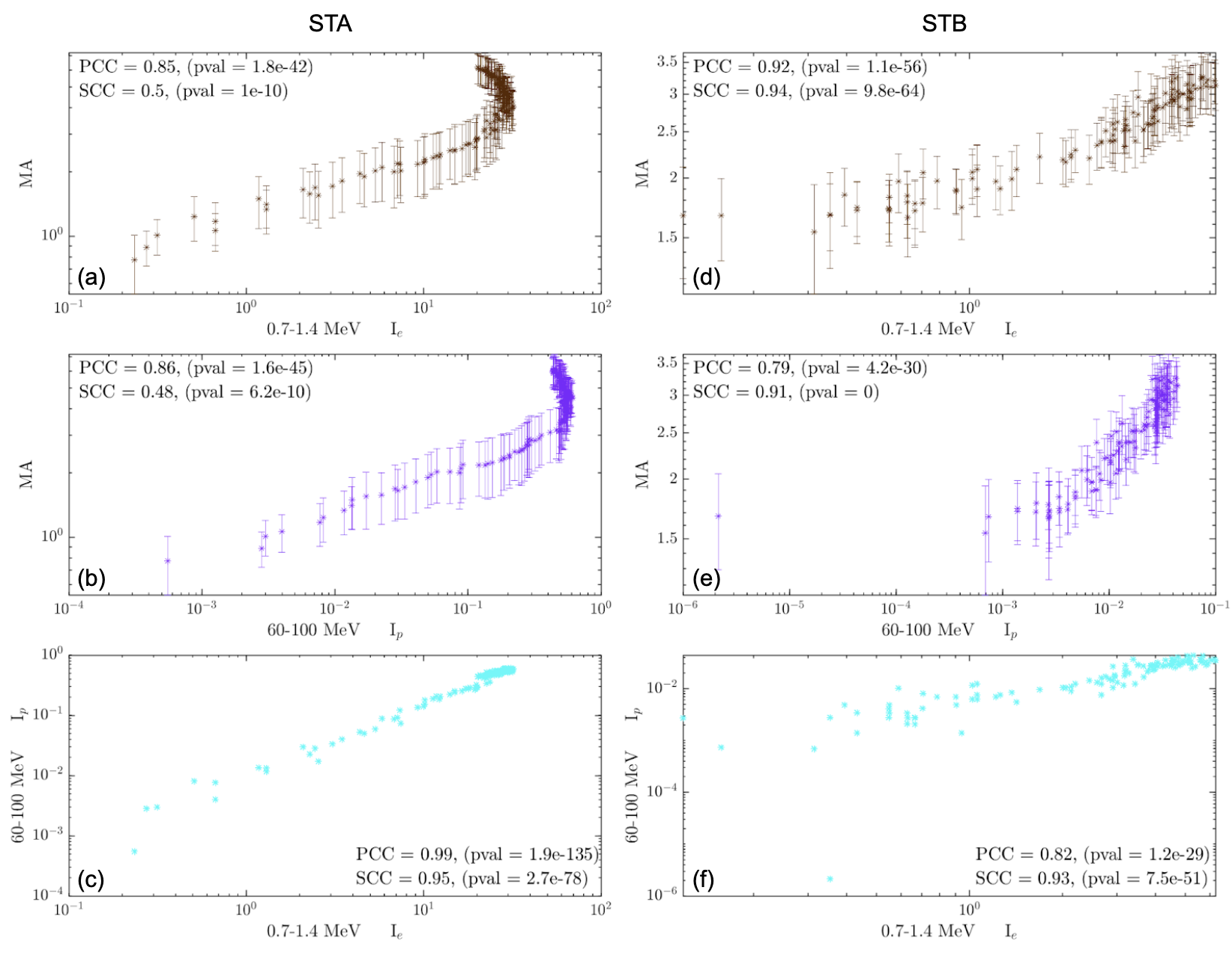}
        \caption{Scatter plots for the calculation of correlations coefficients. Left panels: STA, right panels: STB. Panels (a) and (d): $M_A$ as a function of the electron intensity $I_e$ in the HET energy channel 0.7-1.4 MeV. Panels (b) and (e): $M_A$ as a function of the proton intensity $I_p$ in the HET energy channel 60-100 MeV. Panels (c) and (f): Proton intensity $I_p$ as a function of the electron intensity $I_e$.}
    \label{scatter_plots}
\end{figure*}

Fig. \ref{MA_time-series} shows in common solar time the SEP intensities and the modelled shock parameters (Mach number and shock compression ratios) extracted along field lines connected to STA (left panels) and STB (right panels). The top two rows display modelled shock parameters while the bottom four rows show SEP properties. In all plots the thick lines show nominal values and the dotted lines bound the range of possible shock parameters around the nominal values deduced from the mapping uncertainty. For the SEPs, the dotted lines map the uncertainties provided by the VDA technique.

Orange vertical dashed lines mark the onset time (06:58 UTC minus 3.2 minutes in the Fig.) of the hard X-ray emission deduced from the soft X-ray observations made by the Solar Assembly for X-rays (SAX) on MESSENGER \citep{Schlemm_2007} which directly observed the soft X-rays from the flare site \citet{Plotnikov_2017}. This is followed 12 minutes later by the onset of the type II radio burst (purple dashed line, 07:10 UT minus 8.33 minutes in the Fig., also provided by \citet{Plotnikov_2017}) which confirms that the pressure wave has steepened into a shock, in agreement with the existence of multiple locations on the pressure wave where the Alfvén Mach number exceeds 1.

Overall the time-shifted onsets of electrons (Fig. \ref{MA_time-series}ci) and protons (Fig. \ref{MA_time-series}dj) deduced from SEP measurements coincide well with the pressure wave connection times (first vertical magenta dashed line). The modelling reveals that at these times the wave has steepened into a shock in the relevant regions with compression ratios and Mach numbers exceeding 1 as seen in the top two rows of Fig. \ref{MA_time-series}. We find that the shock is stronger along field lines connected to STA than STB which could explain why the SEP event was strongest at STA around the onset of the event. The magnetic connections of the pressure wave to STA and STB occur at 07:20 UT and 07:29 UT respectively, which is 20 and 30 minutes after the onset of the hard X-rays and 10 and 20 minutes after the shock has formed. \\

The anisotropies of 85--125\,keV electrons (Fig. \ref{MA_time-series}fl) are calculated with the four-telescope STEREO SEPT data using the weighted-sum method of \citet{Bruedern_2018}. We also show the anisotropies for background-subtracted data from which the mean intensity from a steady background window 02:00--07:00 UT has been removed. The results show that both spacecraft are measuring highly anisotropic electron fluxes indicative of a direct magnetic connection to the particle accelerator. Finally we note that both the SEP events at STA and STB are electron rich which could be indicative of either connection to a quasi-perpendicular shock wave \citep[see ][]{Jebaraj_2023} or perhaps these are flare-accelerated particles. An interesting observation is apparent when comparing the results of Fig. \ref{MA_time-series}ci and Fig. \ref{MA_time-series}dj. The proton onsets seem to coincide exactly with the connectivity to the shock (i.e 07:12 for STA and 07:21 for STB in common solar time (ST)), whereas, the electron onsets occurred a few minutes before the shock connection. This could be indicative of flaring electrons diffusing sooner from the flaring site onto the connecting field lines.

\begin{figure*}[t]
    \centering
    \includegraphics[scale=0.5]{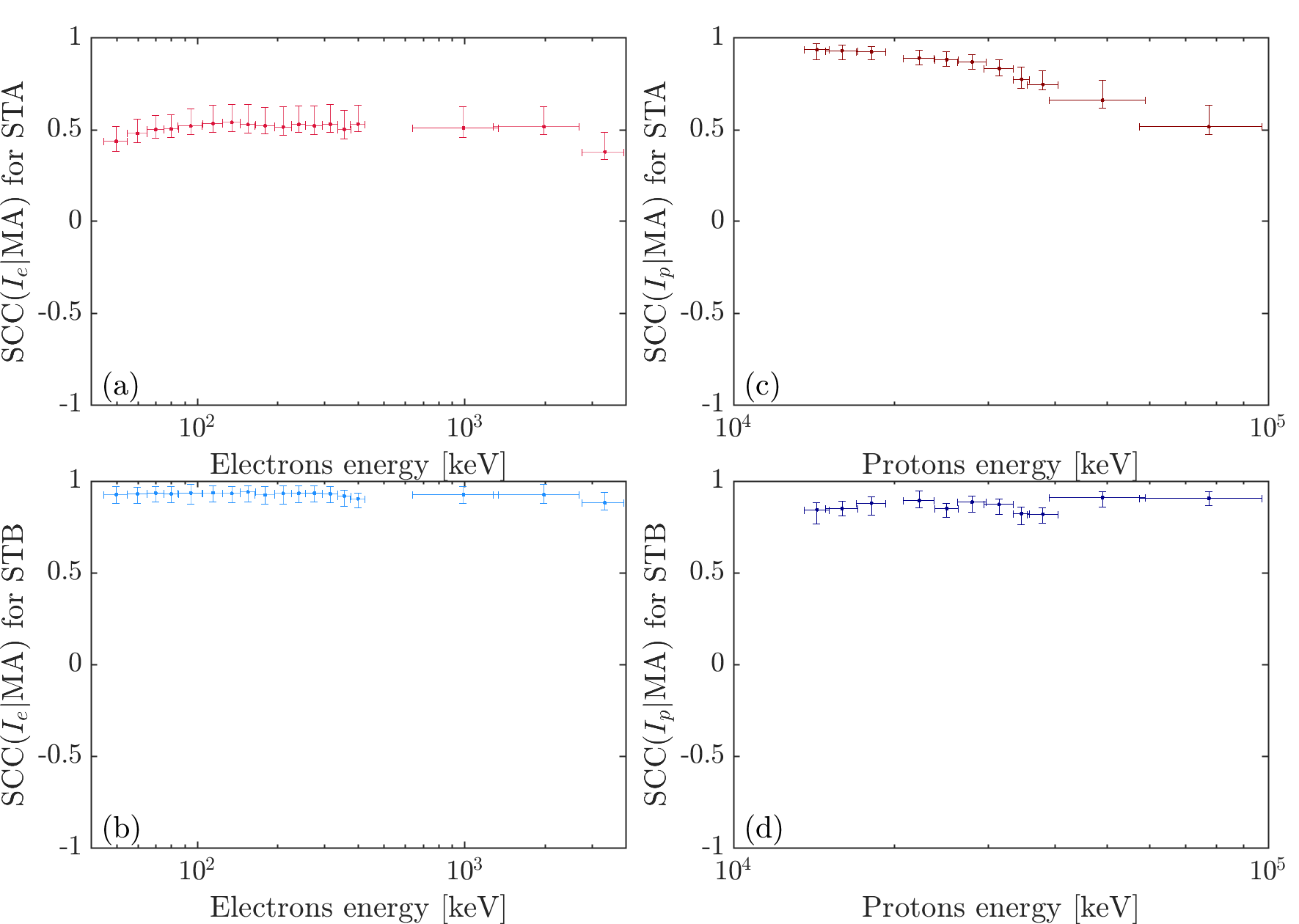}
        \caption{The Spearman's Correlation Coefficient (SCC) between Alfvénic Mach Number (MA) and particle time-series ($I_e$ for electron intensity and $I_p$ for proton intensity) as a function of the particle energy. Panel (a): SCC between log($I_e$) and log(MA) as a function of electron energy for STA. Panel(b): same as panel (a) but for STB.
        Panel (c): SCC between log($I_p$) and log(MA) as a function of proton energy for STA. Panel (d): same as panel (c) but for STB.
        In the four panels, errorbars are defined by the range of SCC values obtained for the hundred magnetic field lines created using MHD cubes. For more details, please see the text.}
    \label{scc_STA_and_STB_MA_e_and_p}
\end{figure*}

\section{Correlation between shock properties and SEPs}

The synchronicity between the onset of SEPs and the times of magnetic connectivity to the shock are indicative of a close connection between the shock evolution and the particle events at the spacecraft. In order to explore this relation further, we examined relations between the time evolving shock parameters and the time intensity of electrons and protons, for each energy channel independently. For that, we select a time range, which starts when the shock connects to the different spacecraft, marked by the pink dashed vertical lines on Fig. \ref{MA_time-series} and ends when the anisotropies (\ref{MA_time-series}fl) have significantly diminished. The level of anisotropy relates to the relative fluxes of freshly accelerated particles and back-scattered particles that have undergone significant diffusion at and beyond the spacecraft location. By limiting this investigation during the times of highly anisotropic particle fluxes we focus on the relation between the accelerator and the early accelerated particles. Diffusive shock acceleration theory implies a close connection between the varying Mach number of the shock and the time-varying intensity of the SEPs \citep{Afanasiev_2018}.

We use two different methods to calculate the correlation coefficients shown in Fig. \ref{scatter_plots}. The Pearson's correlation coefficient (PCC) assesses the linear dependence between two continuous variables and is defined as the covariance of the two variables divided by the product of their standard deviations. Spearman's correlation coefficient (SCC) measures the monotonic relationship between two continuous or ordinal variables. The latter is useful when the variables tend to evolve together, but not necessarily at a constant rate, unlike the PCC where the variables change proportionally. A p-value was calculated for each of these correlation coefficients, and if the latter was greater than 5\%, we automatically removed the calculated coefficient for its lack of statistical significance. Looking at the recorded SEP events, we see that there was no significant intensity received by STA under 500 keV for protons, and STB did not detect significant ion intensities with the SEPT instrument (no clear onset is seen in any of the energy bands). We therefore removed these energy channels from the correlation analysis as they did not bring additional information.

Fig. \ref{scatter_plots} shows scatter plots of $M_A$ (top panel of Fig. \ref{MA_time-series}) as a function of electrons (panels (a) and (d)) and protons (panels (b) and (e)), for STA (left panels) and STB (right panels) and specific energy channels. Two parts can be distinguished in the relation, especially for STA. The first one is quasi-linear, and corresponds to the rising part of the SEP event (from a few tens of minutes to a few hours, depending on energy and species) until the SEP peak. The second part breaks away from this quasi-linearity, and corresponds to the 'plateau' phase of the SEP event after its peak, whereas the $M_A$ continue to grow. The particles continue to be injected (as can be seen from the anisotropy in Fig. \ref{MA_time-series}fl) but diffusion takes over the shape of the SEP.

Fig. \ref{scc_STA_and_STB_MA_e_and_p} presents the combined results for all energies for the Alfvénic Mach Number. For each panel, SCCs between log($M_A$) and particle intensity are given as a function of the particle energy for one spacecraft.
Correlation coefficients are only plotted in the Fig. if they are statistically significant, i.e. with a p-value lower than $0.05$. For the error bars, we calculate the SCC between log($M_A$) and SEP intensities for different $M_A$ values extracted along one hundred magnetic field lines created with the MAST cubes (see Sect. \ref{sect_shock_modelling} for more details).
We can see higher CCs for STB protons than STA, with an apparent energy dependence for STA protons, which is not visible in STB. The correlation coefficients for electrons are high at both spacecraft and don't show a clear energy dependence. 

A similar correlation analysis was performed between SEPs and several other shock parameters, i.e. the shock speed $V_{sh}$, the Hoffman-Teller speed $V_{HT} = V_{sh}/\cos (\theta_{BN})$, its cosine $\cos (\theta_{BN})$, the Magneto-Sonic Mach Number $M_{\rm ms}$, the angle between the shock normal and the magnetic field $\theta_{BN}$, the Hoffman-Teller Mach Number $M_{HT} = M_A/\cos (\theta_{BN})$, and the density compression ratio of the shock (XR). For the Mach numbers we take the logarithms. The results are summarised in table \ref{tab_SCC} which gives Spearman's correlation coefficients between these shock parameters and particle intensities recorded by STA and STB. We choose to give only SCC because the PCC were not significant for some parameters and energies, and SCC appears to reflect better the relation between SEP event profile and shock parameters evolution probably because the shock parameters and SEPs do not evolve proportionally.

For electrons we select two energy bands of the SEPT instrument (45-55 keV and 375-425 keV) and two others from the HET instrument (0.7-1.4 MeV and 2.8-4 MeV). For protons we only select three energy bands of HET (13.6-15.1 MeV, 26.3-29.7 MeV and 60-100 MeV) because SEPT did show a significant ion intensity increase, as explain in Sect. \ref{sect_analyse_particles}.

\begin{table*}[t]
\caption[]{Spearman's Correlation Coefficients (SCC) between the SEP times-series of STA and STB for different energy channels and the shock parameters. In the table all the SCC are significant, i.e their associated p-values are < 0.05.}
\footnotesize
\centering
\begin{tabular}{ c  |  c  c  c  c  |  c  c  c  }
\hline \hline ~SCC~ &  & electron intensity &  &  &  & proton intensity & \\ \hline
STA & 45-55 keV & 375-425 keV & 0.7-1.4 MeV & 2.8-4 MeV & 13.6-15.1 MeV & 26.3-29.7 MeV & 60-100 MeV \\ \hline
$V_{sh}$ & 0.52 & 0.6 & 0.6 & 0.46 &  {\color{red}0.88} & 0.85 & 0.59\\
$V_{HT}$ & 0.24 & 0.32 & 0.31 & 0.16 &  {\color{red}0.66} & 0.63 & 0.29\\
$M_A$ & 0.44 & 0.52 & 0.5 & 0.37 &  {\color{red}0.93} & 0.86 & 0.48 \\
$M_\mathrm{ms}$ & 0.46 & 0.55 & 0.52 & 0.4 &  {\color{red}0.9} & 0.84 & 0.5\\
$M_{HT}$ & 0.4 & 0.48 & 0.46 & 0.34 &  {\color{red}0.87} & 0.81 & 0.45\\
$\theta_{BN}$ & -0.37 & -0.35 & -0.38 & -0.37 & -0.25 & -0.25 &  {\color{red}-0.42} \\
XR & 0.45 & 0.53 & 0.5 & 0.38 &  {\color{red}0.92} & 0.85 & 0.49\\ \hline
STB & 45-55 keV & 375-425 keV & 0.7-1.4 MeV & 2.8-4 MeV & 13.6-15.1 MeV & 26.3-29.7 MeV & 60-100 MeV \\ \hline
$V_{sh}$  & 0.83 & 0.83 &  {\color{red}0.87} & 0.83 & 0.74 & 0.79 & 0.84\\
$V_{HT}$ & 0.25 & 0.31 & {\color{red}0.32} & 0.29 & 0.23 & 0.24 & 0.28 \\
$M_A$  & 0.93 & 0.92 &  {\color{red}0.94} & 0.89 & 0.86 & 0.89 & 0.91\\
$M_\mathrm{ms}$ & 0.92 & 0.91 &  {\color{red}0.94} & 0.89 & 0.85 & 0.88 & 0.91\\
$M_{HT}$ & 0.75 & 0.77 &  {\color{red}0.77} & 0.72 & 0.71 & 0.73 & 0.74 \\
$\theta_{BN}$ & -0.89 & -0.87 &  {\color{red}-0.92} & -0.88 & -0.8 & -0.83 & -0.89\\
XR & 0.92 & 0.91 &  {\color{red}0.94} & 0.89 & 0.85 & 0.89 & 0.91\\ \hline
\end{tabular}
\label{tab_SCC}
\end{table*}

In table \ref{tab_SCC} we write in red the highest SCC value for each shock parameter. STA shock parameters correlate less with electron profiles of all selected energies compared to STB. For protons, and especially for STA, correlations are more variable according to energy for a given shock parameter.

For protons, the highest correlation coefficients are found overall with the shock speed, magneto-sonic, Alfvén Mach number ($M_A$) and the shock compression ratio. All these parameters are closely related to the speed of the shock flanks and the coronal conditions. For the highest energy protons measured at STA (60-100~MeV), the correlation coefficients were significantly lower than for lower energies suggesting that additional processes are in play. At STB, the correlations remain very high for all energy bands.

For electrons, the correlation coefficients between all shock parameters (except $V_{HT}$) are very high at STB and significantly lower at STA suggesting a stronger link between shock evolution and particle fluxes along STB field lines and particle acceleration.

The shock geometry $\theta_{BN}$ is also more weakly anti-correlated with particle intensities measured by STA than at STB with a tendency for electrons to anti-correlate more significantly than protons.

Summarising the above results, we find that that overall the time-varying flux of electrons and protons are correlated with shock parameters, especially along field lines connected to STB. For STA, electrons, as well as, the highest energy protons exhibit lower correlations suggesting that at this observer additional processes may have contributed to the acceleration of the observed SEPs.

\section{Composition of the SEPs}
\label{sec: composition}

\begin{figure}[t]
\centering
    \includegraphics[scale=0.47]{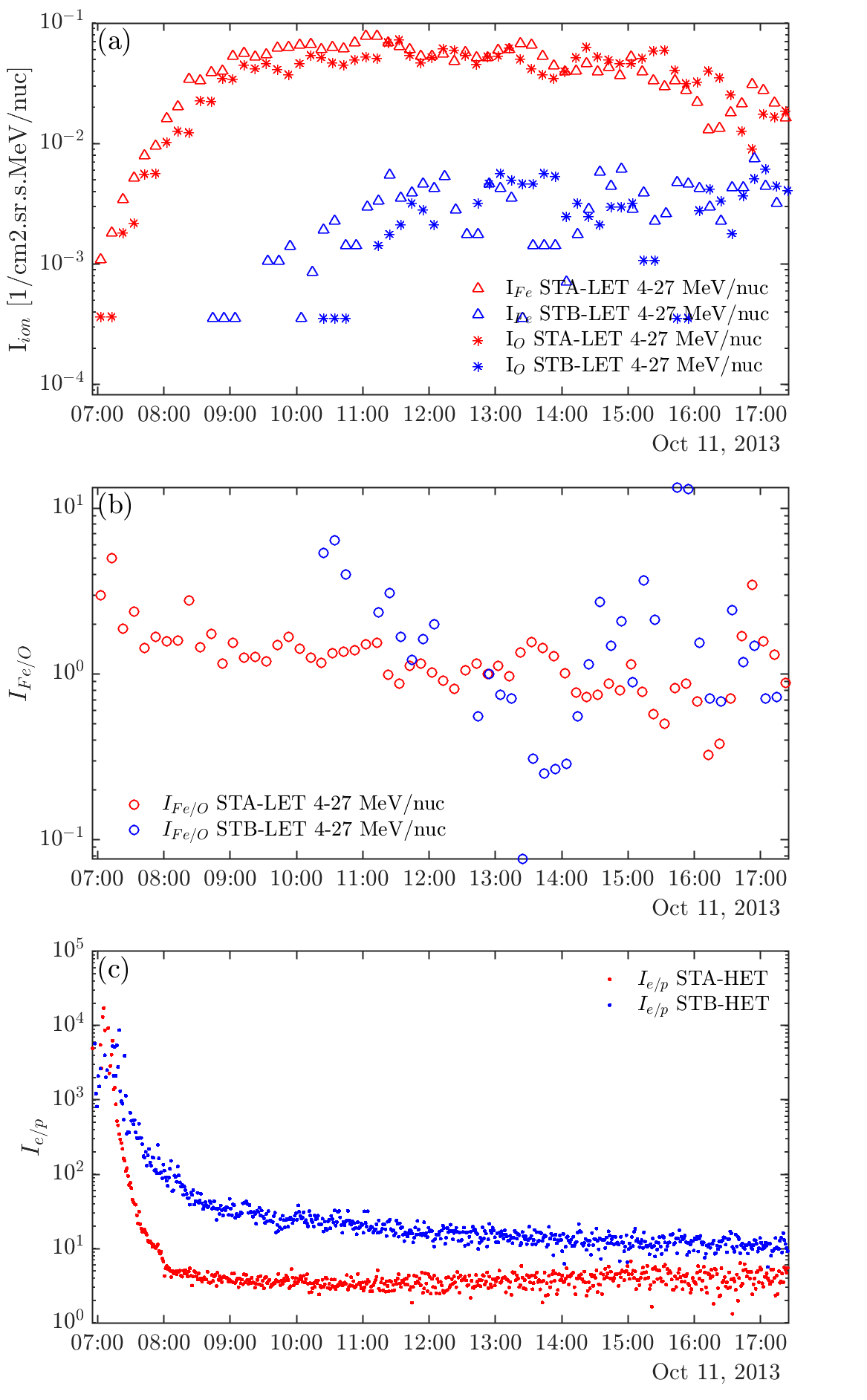}
        \caption{Panel (a): Iron and oxygen intensity (respectively $I_{Fe}$ in triangles and $I_{O}$ in stars) as a function of time, in red from STA/LET instrument and in blue data from STB/LET instrument. Intensities are time shifted according to the VDA.
        Panel (b): Iron to oxygen ratio ($I_{Fe/O}$) as a function of time, in red for STA and blue for STB, time shifted according to the VDA.
        Panel (c): 2.8-4 MeV electron to 60-100 MeV proton ratio as a function of time, time shifted according to the VDA (before calculation of the ratio), in red for STA and blue for STB.}
    \label{STA_and_STB_FeO_and_ep}
\end{figure}

Additional insights on this SEP event are provided by the composition measurements made by the LET instrument. The top panel of Fig. \ref{STA_and_STB_FeO_and_ep} presents the time-varying Fe and O fluxes measured at STA (red) and STB (blue) in the 4-27 MeV/nuc range. The onset of energetic protons at STA is associated with simultaneous increase in both Fe and O fluxes. This is in stark contrast to STB that did not measure significant fluxes of Fe and O until after 11:00 UT, well after the proton onset and despite being magnetically connected to the shock already before 08:00UT.\\

The middle panel of the same Fig. shows that at STA the Fe/O ratio is elevated at proton onset with values exceeding the typical threshold assigned to flare-related events \citep[> 0.134, the coronal value]{Reames_1999}. There is also a tendency for the Fe/O ratio to decrease with time at STA while the intensity of Fe and O fluxes (top panel) remain elevated. At STB the Fe/O ratio is also elevated but more variable.

\begin{figure}[t]
\centering
    \includegraphics[scale=0.45]{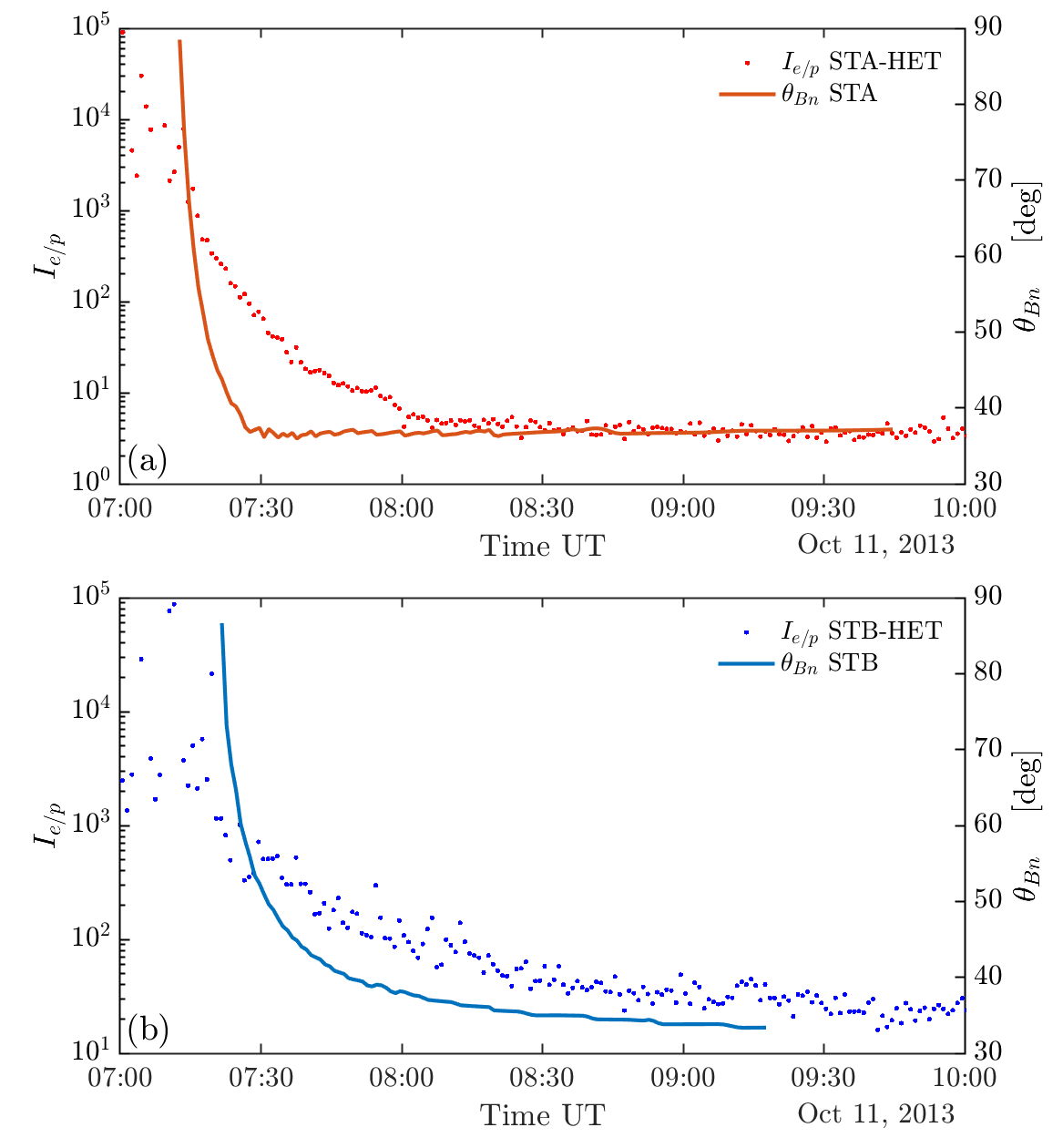}
        \caption{$I_{e/p}$ (left axis) and $\theta_{Bn}$ [deg] (right axis) as a function of time, for STA (red, panel a) and STB (blue, panel b), time shifted according to VDA and light travel time.}
    \label{fig_ep_and_thBN_sta_and_stb}
\end{figure}

The bottom panel of Fig. \ref{STA_and_STB_FeO_and_ep} presents the electron to proton (e/p) ratio as a function of time. Before calculating the e/p ratio (at the following energy ranges 2.7-4 MeV and 60-100 MeV for electrons and protons respectively) we shifted the curves to the solar time (ST) according to their travel time calculated with the VDA. The maximum of this ratio is one order of magnitude higher for STA and decreases very rapidly with time compared to STB, suggesting a efficient and short-lasting electron acceleration. Fig.  \ref{fig_ep_and_thBN_sta_and_stb} shows that the e/p ratio is closely related in time to the quasi-perpendicular phases of the shock at the two spacecraft. Quasi-perpendicular shocks are expected to be highly efficient accelerators of electrons through shock-drift acceleration. 

We conclude that the acceleration conditions and the processes involved near the footpoint of magnetic field lines of STA and STB were different with an enrichment of electrons at STA. We will discuss in the Sect. \ref{sect_interpretation} the possible pre-conditioning of the shock-acceleration process by the flare that occurred before the shock passage.

\section{Discussion}
\label{sect_interpretation}

In this paper we present a methodology to study temporal relations between the time profiles of shock wave parameters and SEP intensities of the associated SEP events, with the aim to better understand the acceleration processes that affect the time-evolution of SEP fluxes measured in situ. We chose one event in the catalogue of triangulated shock waves of \citet{Kouloumvakos_2019}, the 11 October 2013, which presents a difference in the observed SEP intensities \citep{Klassen_2016} between two well separated spacecraft. We focused our study on the link between the onset of a powerful SEP event and the expansion of a pressure wave expanding in the solar corona during the first few minutes of the eruption.

Overall we have found a remarkable synchronicity between the onset of the energetic particles at STA and STB and the time of magnetic connectivity of these spacecraft with the expanding shock (Fig. \ref{MA_time-series}). Furthermore the relative intensity of the SEPs at the two spacecraft could be related to some extent to the relative strength of the shock wave along the relevant magnetic field lines (Fig. \ref{MA_time-series}).

Focusing on the time window during which the particle fluxes are highly anisotropic which corresponds to when the spacecraft are directly connected to the particle accelerator and particles have not yet started to significantly back-scatter during their transport (Fig. \ref{MA_time-series}), we have looked at correlations between the temporal evolution of shock parameters and particle fluxes. The results reveal an overall high correlation with the shock parameters, for both protons and electrons, that suggests an important role of the shock in accelerating both species (Fig. \ref{scatter_plots}). This supports previous statistical studies that focused on a global timescale for several SEP events and showed that there is a significant correlation between the maximum Alfvén Mach number of shock waves and the maximum of the SEP fluxes \citep{Kouloumvakos_2019, Dresing_2022}.\\

Zooming out on the daily timescale of the SEP event, the SEP profiles of STA and STB present clear differences as we showed in Fig. \ref{STA_and_STB_SEP} and discussed in Sect. \ref{sect_event_presentation}. The more flatter shape of the SEP intensity profiles at STB can be easily explained in terms of the direction of the shock (Fig. \ref{Solar_Mach}), which eventually hits STB on 12 October around 18 UT implying that the shock keeps on accelerating particles during its propagation \citep{Lario_2018}. This is also verified with overall longer-lasting particle flux anisotropies measured by STB (Fig. \ref{MA_time-series}), suggesting indeed that new particles are continuously accelerated and reach the spacecraft. In an on-going study, we parameterise particle production at the shock and model these transport processes for the entire event.

Of particular interest in the present analysis are aspects of the event that are not easily explained in terms of particle acceleration at the shock. First, the SEP event measured at STA was much stronger than at STB, with a shock strength higher along STA field lines. However, the shock was sub-critical (Mach less than 2.7) at the connection with STA field lines and not much stronger than at the connection with STB (Fig. \ref{MA_time-series}). The correlation between electrons and $M_A$ at STA is not strong, compared to that between electrons and $M_A$ at STB (Table \ref{tab_SCC}).
This can also be seen in the correlation between SEPs and $\theta_{BN}$. Second, STA measured a high flux of heavy particles right at the onset of the SEP event while STB did not (panels (a) and (b) of Fig. \ref{STA_and_STB_FeO_and_ep}). This leads us to think that an additional process contributed to the acceleration of SEPs along STA field lines (Fig. \ref{STA_and_STB_SEP}). 

\begin{figure}[t]
    \centering
    \includegraphics[scale=0.45]{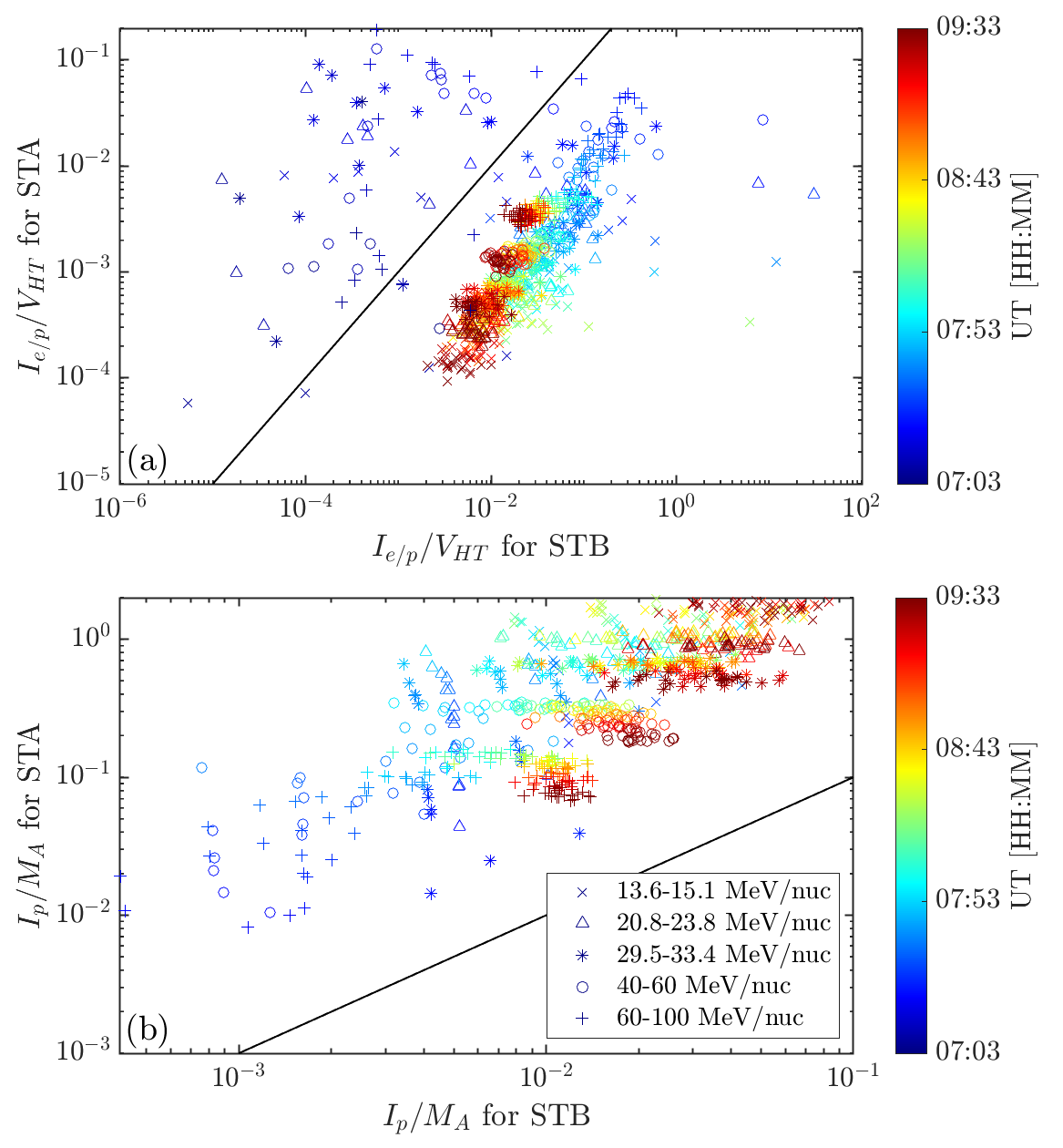}
        \caption{Panel (a): $I_{e/p}$ to $V_{HT}$ ratio for STA as a function of $I_{e/p}$ to $V_{HT}$ ratio for STB. $I_{e/p}$ is the ratio of electron intensity ($I_e$) to proton intensity ($I_p$). The energy band for electrons is fixed to 2.7-4 MeV and the energy band for protons is 13.6-15.1 MeV (crosses), 20.8-23.8 (triangles), 29.5-33.4 MeV (stars), 40-60 MeV (circles), and 60-100 MeV (plus). $I_e$ and $I_p$ are time-shifted according to the VDA before calculation of their ratio. The black line represents the case where these to ratios are equals.
        Panel (b): $I_p$ to $M_A$ ratio for STA as a function of $I_p$ to $M_A$ ratio for STB. The proton intensities are the same as described in panel (a). The black line represents the case where these to ratios are equals.}
    \label{fig_ratio_shock_param_STA_vs_STB}
\end{figure}

We also find that the e/p ratio calculated at STA is very high and impulsive at the onset of the SEP event, whereas it is lower and more gradual at STB (panel (c) of Fig. \ref{STA_and_STB_FeO_and_ep}). For both spacecraft the evolution of the ratio is correlated with the evolution of the shock geometry (Fig. \ref{fig_ep_and_thBN_sta_and_stb}).
Panel (a) of Fig. \ref{fig_ratio_shock_param_STA_vs_STB} show the $I_e/I_p$ to $V_{HT}$ ratio at STA as a function of that same ratio at STB, for five different proton energy bands and a fixed electron energy band of 2.7-4 MeV (the highest of HET). We clearly see, for equivalent $V_{HT}$, a higher flux of electrons compared to protons for STA during the first minutes of the event and the tendency is inverted after approximately 20 minutes. The excess electrons during the early phase are likely related to the shorter but highly quasi-perpendicular shock geometry at STA (Fig. \ref{fig_ep_and_thBN_sta_and_stb}). This is because electron acceleration at shock waves is particularly efficient at quasi-perpendicular shocks through shock-drift acceleration. However Fig. \ref{fig_ratio_shock_param_STA_vs_STB} suggests
that enhanced electron fluxes at STA during the very early phase of the event could result from additional effects such as electron suprathermals produced during a preceding solar flare at the footpoint of STA magnetic field line.\\
The excess particle fluxes at STA is most pronounced and of longer duration (several hours) in the proton channels. The theory of proton acceleration at shock waves through diffusive-shock acceleration gives a strong relation between shock strength that can be be quantified through the shock Mach number. In panel (b) of the same Fig., we compare the $I_p$ to $M_A$ ratio for STA with the one of STB. During the first hours of the event, for equal $M_A$, STA measured significantly higher fluxes of protons than STB, and this is visible whatever the protons energy.\newline

We conclude that the early phase of the SEP event measured by STA is significantly stronger and enriched in particles that are typically accelerated in flares. We suggest that are pre-event flaring activity 3 hours before the SEP event enhanced the population of suprathermal particles close to STA's magnetic footpoint location. This flare can be seen around 03:35 UTC on the same day in EUV wavelengths thanks to EUVI instrument onboard STA and STB (Fig. \ref{magnetogram}). As discussed in Sect. \ref{sect_event_presentation}, this flare was lacking any associated CME and type III radio bursts suggesting confined conditions, but would have nevertheless produced a population of energetic ions trapped in the lower corona. This population could have served as seed particles enriching the SEP event produced by the main eruption of 07:25 UTC. This could explain the higher peak intensity observed for protons, electrons and ions recorded by STA and its impulsive shape, since the shock only cannot.\\

\begin{figure}[t]
    \centering
    \includegraphics[scale=0.2]{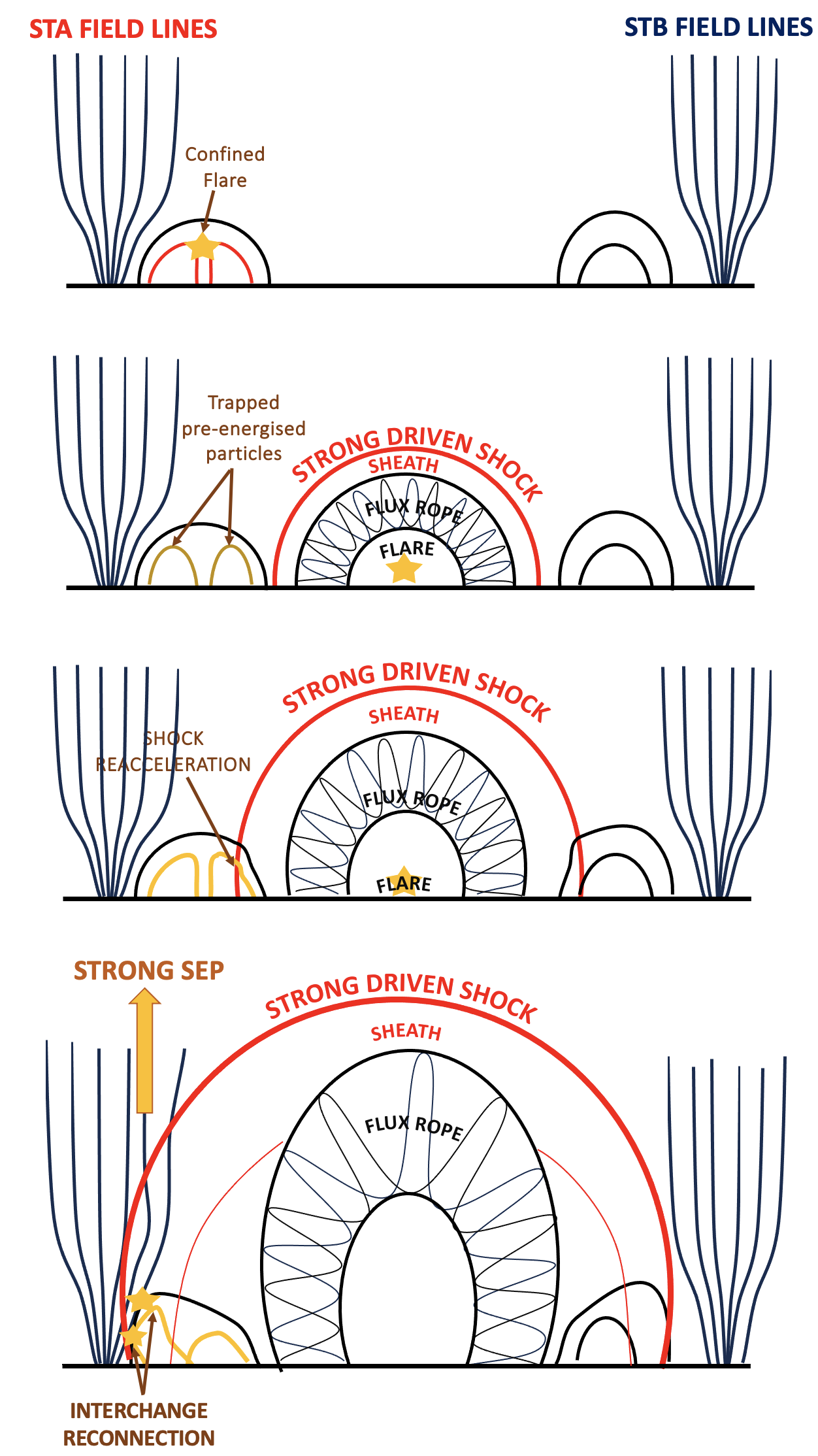}
        \caption{Proposed sequence of events to explain the properties of the 11 Oct 2013 SEP intensities at STA and STB. The top panel represents the configuration of magnetic field lines before the event, with STA-connected field lines on the left and STB-connected field lines on the right. Red lines on top panel represent the confined flare of $\approx$ 03:35 UT. As the pressure wave forms around the expanding CME flux rope, open and closed magnetic field lines are pushed aside and potentially forced to undergo interchange reconnection. Initially trapped pre-energised particles are suddenly released on open fields into the vicinity of the shock for further acceleration.}
    \label{scheme_event}
\end{figure}

We therefore propose the following concomitant sequence of energisation processes schematised in Fig. \ref{scheme_event}. A flare occurring three hours before our event pre-accelerated particles in the region of magnetic connectivity of STA field lines providing a trapped particle population of pre-accelerated particles (top of Fig. \ref{scheme_event}). As the pressure wave driven by the expanding CME expands, the corona is disturbed over an increasingly large area. The sheath behind the forming shock forces field lines to interact and potentially undergo magnetic reconnection. 

Any pre-energised particles that are trapped in magnetic loops and that have not yet precipitated into the chromosphere can suddenly be released along open field lines that have interchanged with closed loops. These particles, now free to diffuse from the sheath downstream into the shock region, are then accelerated in an initially highly quasi-perpendicular shock to potentially much higher energies \citep{Tylka_2005, Sandroos_2009}. The composition of the event then reflects that of the preceding flare due to the energy threshold at which particle acceleration becomes efficient for quasi-perpendicular shocks according to the theory of \citet{Tylka_2006}.

\section{Conclusion and Perspectives}
\label{sect_discussion}

We plan to apply our new methodology to a larger number of events, for example those triangulated shock waves from the catalogue of \citet{Kouloumvakos_2019}, will allow us to uncover similarities and regular patterns in the connection between shock parameter and SEP intensity-time series around onset. In addition we could compare events where any preceding flaring has occurred at the point of magnetic connectivity of one spacecraft but not the others. We are carrying out a search of such configurations during times covered by the more recent Parker Solar Probe and Solar Orbiter spacecraft.

Finally, as already mentioned, we are in the process of modelling this event and others in the list of  \citet{Kouloumvakos_2019} by parameterising particle production at the shock in terms of shock parameters and propagating these particles using a particle transport code \citep{vandenBerg_2020, Strauss_2023}. This will provide some quantification of the excess particle fluxes along certain magnetic field lines that might be related to pre-conditioning effects.

\begin{acknowledgements}
    This project has received funding from the European Union’s Horizon 2020 research and innovation program under grant agreement No 101004159 (SERPENTINE project, \url{https://serpentine-h2020.eu/}). This study has been partially supported through the grant EUR TESS N°ANR-18-EURE-0018 in the framework of the Programme des Investissements d'Avenir. N.D. and L.V. are grateful for support by the Research Council of Finland (SHOCKSEE, grant No.\ 346902). A.K. acknowledges financial support from NASA HGIO grant 80NSSC24K0555.
\end{acknowledgements}

\bibliographystyle{aa} 
\bibliography{main.bib} 

\end{document}